# Fast Transfer-free Synthesis of High-quality Monolayer Graphene on Insulating Substrates by Simple Rapid Thermal Treatment

*Zefei Wu,[a)] Yanqing Guo,[a)] Yuzheng Guo,[a)] Rui Huang,[*] Jie Song, Zhenxu Lin, Huanhuan Lu, Jiangxiazi Lin, Shuigang Xu, Yu Han, Hongliang Li, Yuan Cai, Chun Cheng, Dangsheng Su, John Robertson, and Ning Wang[*]*

[a)] These authors contributed equally to this paper.

Zefei Wu, Huanhuan Lu, Jiangxiazi Lin, Shuigang Xu, Yu Han, Yuan Cai, Ning Wang

Department of Physics, the Hong Kong University of Science and Technology, Hong Kong, China

E-mail: phwang@ust.hk

Yanqing Guo, Rui Huang, Jie Song, Zhenxu Lin, Hongliang Li

Department of Physics and Electronic Engineering, Hanshan Normal University, Chaozhou, China

E-mail: rh628@cam.ac.uk

Yuzheng Guo, Rui Huang, John Robertson

Department of Engineering, University of Cambridge, Cambridge, United Kingdom

E-mail: rh628@cam.ac.uk

Chun Cheng

Department of Materials Science and Engineering, South University of Science and Technology of China, Shenzhen, China

Dangsheng Su

Shenyang National Laboratory for Materials Science, Institute of Metal Research, Chinese Academy of Sciences, Shenyang, China

Graphene, a quasi-2D material, has attracted considerable attention because of its excellent transport properties, which make it a promising material for applications ranging from radio-frequency devices and transistors to optoelectronic devices.[1–5] The tremendous interest in graphene has led to various kinds of preparation methods,[6–17] such as exfoliation from bulk graphite,[9] chemical vapor deposition (CVD) on transition metals,[10–16] and reduction of graphite oxide.[17] Among these, CVD on transition metals appears to be the most promising for producing high-quality, large-scale graphene; in this method, carbon sources, such as methane and acetylene, are commonly used as precursors with metal catalytic substrates.[10–13, 18–19] Although the CVD growth method possesses advantages such as low cost and large scale, using metal catalytic substrates requires a process for transferring graphene onto a desired substrate for further applications. Such transfer process is not only inconvenient but also causes additional contamination, wrinkling, and breakage of the graphene, which result in problems in the devices. To overcome these issues, the transfer-free synthesis of graphene on a desired substrate has been attempted using two main strategies.[20–32] The first approach involves directly growing graphene by metal-catalyst-free CVD on dielectric substrates such as $SiO_2$,[20–22] $Al_2O_3$,[23–24] BN,[25–27] and $SrTiO_3$.[28] However, producing high-quality, large-scale graphene using this method is difficult. The second approach involves directly synthesizing graphene by thermally converting a solid carbon film coated on insulating substrates through a metal catalyst capping layer.[29–31] Recently, Zhuo *et al*. reported that the carrier mobility of graphene produced by this method could reach up to 1835 $cm^2$ $V^{−1}s^{−1}$.[32] Apart from these two main strategies, the transfer-free synthesis of graphene is also performed through metal-vapor-assisted CVD, in which a metal catalyst in vapor form reacts with carbon precursor gases in the gas phase as well as on the substrate surface.[33–34] This process enables metal residue-free growth of high-quality, large-scale graphene comparable with graphene grown on metal foil in terms of structural defect level. However, the as-grown graphene exhibits a maximum carrier mobility of only ~800 $cm^2V^{-1}s^{-1}$ at room temperature.[34] Therefore, the development of large-scale graphene with high quality through a simple process remains a considerable challenge.

For CVD-grown graphene, nickel and copper are the commonly used metal catalysts. However, the growth kinetics and mechanism between nickel and copper are different because of the varying solubilities of carbon in these metals.[10–11, 30] In the case of nickel, where carbon solubility is high, graphene growth is governed by segregation and precipitation processes, which generally produce multilayer graphene.[30] To suppress multilayer formation, Weatherup et al. introduced a carbon diffusion barrier ($Al_2O_3$) inserted into an amorphous-C/Ni bilayer stack, which effectively prevented premature carbon dissolution and significantly improved monolayer graphene (MLG) formation.[35] In the current work, the extremely low carbon solubility in copper is considered by inserting a copper film between a solid carbon layer and $SiO_2$. We then demonstrate a simple rapid thermal treatment (RTT) method for the fast and direct growth of large-scale MLG on a $SiO_2$/Si substrate from solid carbon sources. During the RTT process, the copper film inserted between the solid carbon layer and the $SiO_2$/Si substrate does not only act as an active catalyst for the carbon precursor but also as a "filter" that prevents premature carbon dissolution, and thus, contributes to the formation of MLG on $SiO_2$. The produced MLG exhibits high carrier mobility and standard half-integer quantum oscillations. This RTT method is a simple process that cannot only produce high-quality MLG with a scale as large as ~1 $mm^2$ but can also be applied to most catalyst materials. Amorphous carbon (α-C) from carbonized poly(methyl methacrylate) (PMMA) or sucrose ($C_{12}H_{22}O_{11}$), which is used as a solid carbon source in this case, also significantly simplifies the synthesis process of graphene.

**Figure 1**(a) illustrates the substrate preparation for growing MLG on a $SiO_2$/Si wafer though the RTT process. First, a 500 nm-thick copper film was deposited on a silicon wafer covered with 300 nm $SiO_2$ using a DC sputtering system. Then, an annealing process was performed at 1000 °C. Before depositing the solid carbon source onto the copper film, the substrate covered with copper film was immersed in dilute hydrochloric acid to remove copper oxide. Two carbon sources, PMMA and sucrose, were spin-coated on the surface of the copper films. The PMMA- and sucrose-coated samples then underwent RTT, as depicted in Figure 1(b). During this treatment, a sample was rapidly heated to approximately 1000 °C at a heating rate of approximately 160 °C/s. After heating the sample at the maximum temperature for 4–8 minutes, it was then cooled down to room temperature for approximately 30 minutes. During the entire annealing process, the heating chamber was filled with ultrapure argon gas to prevent oxidation. MLG directly formed at the copper–$SiO_2$ interface, whereas α-C remained on the surface of the copper film. This film, along with α-C, was finally dissolved in a 0.1 M $FeCl_3$ solution. Finally, the MLG that formed directly on the $SiO_2$/Si wafer was obtained without any further process.

**Figure 2**(a) shows the typical Raman spectrum of the PMMA-derived as-grown graphene using the RTT method at 1000 °C. The Raman spectrum clearly presents a sharp 2D peak at ~2680 $cm^{-1}$, along with a very weak D peak at ~1350 $cm^{-1}$ and a G peak at ~1580 $cm^{-1}$. The 2D band can be fitted well to a single Lorentzian shape with a full width at half maximum (FWHM) of approximately 40 $cm^{-1}$. Furthermore, the $I_{2D}/I_G$ integrated intensity ratio is ~1.8. These data indicate that graphene is grown in a monolayer. In addition, the $I_D/I_G$ integrated intensity ratio is only as low as ~0.05, which shows the low population of $sp^3$-type defects in graphene. According to the $I_D/I_G$ integrated intensity ratio, the grain size of our graphene is 210 nm as estimated by the empirical relation:

$$L_a(\text{nm}) = 2.4 \times 10^{-10} \lambda^4 (I_D/I_G)^{-1},$$

where $\lambda$ is 514.5 nm.[36–37] The as-grown graphene film was further examined by transmission electron microscopy (TEM). Then, it was detached from the $SiO_2/Si$ substrates using a buffered oxide etch and was suspended on a copper grid for TEM measurements. The clear hexagonal electron diffraction pattern with an inner intensity stronger than the outer intensity, as shown in Figure 2(b), further confirms that the as-grown graphene film has a single layer.[38] The aforementioned results demonstrate that the RTT process produces high-quality MLG on a $SiO_2/Si$ substrate with low degrees of structural defects. Figure 2(c) presents the Raman spectrum measured from the upper surface of a copper film. A high Raman background signal that results from copper surface-plasmon emission forms a broad band with two intense peaks: a 1D peak and a G peak. The as-grown graphene films are different from the graphene films grown on copper foil by regular CVD.[10–11] The excess carbon source supply by PMMA and the short annealing period may be responsible for the deteriorated quality of the graphene films on the upper surface of the copper foil. Therefore, the formation of MLG on a $SiO_2/Si$ substrate indicates that the copper film is not only an active catalyst for carbon precursor but also acts as a "filter" to prevent premature carbon dissolution, which contributes to the formation of MLG on $SiO_2$. Interestingly, the RTT process can generate high-quality MLG as large as ~1 mm$^2$. A uniform color contrast between the MLG region and bare $SiO_2/Si$ in the bright-field optical microscope image shown in Figure 2(d) reveals a large-size MLG on a $SiO_2/Si$ substrate.

One of the advantages of the RTT process is that MLG grown on $SiO_2$/Si can be realized by using different solid carbon sources. The same growth condition using $C_{12}H_{22}O_{11}$ as a carbon source was adopted for the graphene growth on $SiO_2$/Si. **Figure 3** shows the Raman spectrum of the graphene grown on $SiO_2$/Si using $C_{12}H_{22}O_{11}$ as the carbon source. The Raman spectrum exhibits $I_{2D}/I_{1D} \approx 1.8$ with a 2D band that is fitted well to a single Lorentzian shape with an FWHM of ~40 $cm^{-1}$ and featuring a typical characteristic of MLG. The negligible 1D peak further indicates low degrees of structural defects in the MLG. Apparently, $C_{12}H_{22}O_{11}$ can also generate high-quality MLG similar to that obtained using PMMA as the carbon source. As shown in Figure 3, the 1D and 2D bands of the PMMA-/sucrose-derived MLG grown using the RTT process peaked at ~1350 $cm^{-1}$ and ~2680 $cm^{-1}$, respectively. Compared with the MLG grown by regular CVD on copper foil, the PMMA-/sucrose-derived MLG grown using the RTT process exhibits obvious blue shifts in both 1D and 2D bands. These shifts are generally attributed to compressive strain resulting from the different thermal expansion coefficients between graphene and a substrate.[39]

In the RTT growth method, growth temperature plays a crucial role in forming MLG on $SiO_2$/Si. The Raman spectra in **Figure 4** exhibit the relationship between growth temperature and the structural characteristics of the graphene films. The weak 1D, G, and 2D peaks start appearing at a low temperature of 800 °C. Then, these peaks synchronously become intense after the growth temperature increases to 900 °C. The random nucleation and poor surface migration of carbon are presumed to be the major causes of the enhanced 1D peak associated with the defects at these growth temperatures. The defective graphene structure is significantly ameliorated by increasing growth temperature by up to 1000 °C, during which the 1D peak becomes indistinguishable. However, given the growth-temperature dependence of the graphene structure, the carbon atoms that dissociated from PMMA are able to diffuse from the upper surface of the copper film and onto the SiO$_2$/Si surface even at a low temperature of 800 °C. This finding indicates that the grain boundary of the copper film serves as a passage for carbon atoms accessing the SiO$_2$/Si surface during the growth process. To gain a detailed understanding of this behavior, density functional theory is applied to simulate the diffusion pathway of carbon in copper thin film. Cu(111) crystal grains are dominant in the copper thin film, as shown in **Figure 6**(a); thus, the diffusion barriers against carbon in the copper crystal and the grain boundary have been calculated in terms of Cu(111), and the results are shown in **Figure 5**. In this case, the interstitial site with the lowest energy in the copper crystal is identified as an octahedral site, as shown in Figure 5(a). Our calculation indicates that the diffusion barrier from one octahedral site to a neighboring site is 1.1 eV. The grain boundary is simulated by an edge dislocation, as shown in Figure 5(b). Different twinning grain boundaries can be regarded as a series of edge dislocations. The lowest-energy site is located immediately below the extra copper column. The energy barrier for carbon to diffuse along the edge defect line is calculated to be only 0.2 eV, which is significantly lower than the carbon diffusion barrier in the copper crystal. Therefore, we consider the grain boundary as the main diffusion path during graphene growth.

To understand the growth mechanism of graphene further, the effect of the copper film on the graphene structure before and after the annealing process was investigated. Figure 6(a) shows the X-ray diffraction (XRD) spectra of the copper film before (in black) and after (in red) the annealing process. The XRD spectra reveal various crystal orientations of Cu(111), Cu(200), Cu(220), Cu(311), and Cu(222). The diffraction peaks become notably sharp after the annealing process, particularly the (111) diffraction peak, which indicates a remarkable enhancement in crystal fraction in the copper film after the annealing process. Based on the diffraction peaks, the grain sizes of different crystal orientations can also be calculated using the Debye–Scherrer formula, as follows:

$$D = \frac{0.9\lambda}{\beta \cos\theta},$$

where D is the grain diameter size, $\lambda$ is the X-ray wavelength (0.15406 nm), $\beta$ is the FWHM, and $\theta$ is the diffraction angle.[40] Copper grain size before annealing is only as large as 15 nm, using the (111) diffraction as reference. After the annealing process, grain size remarkably increases to 42 nm. The increased grain size and crystal fraction obviously contribute to the growth of high-quality MLG, as indicated in Figure 6(b). In general, the graphene film grown using the copper-catalyzed CVD process is achieved by the surface absorption and decomposition of hydrocarbon precursors through the active copper surface, in which low carbon solubility in copper results in the self-limited growth of MLG.[41–42] The difference between our proposed process and the copper-catalyzed CVD process is that the precursors in our work should be initially diffused from the upper surface of the copper film onto the $SiO_2$/Si surface through the grain boundary, as previously discussed. Given that carbon solubility in copper crystal is low, the grain boundary is regarded to provide passage, through which small doses of the precursors can access the $SiO_2$/Si surface. After reaching the $SiO_2$/Si surface, the precursors will undergo a process similar to that in copper-catalyzed CVD to induce MLG growth. During this stage, grain boundaries significantly influence graphene nucleation and growth.[41, 43–44] Previous experiments revealed that grain boundaries could induce nucleation sites and form continuous non-uniform polycrystalline graphene film with numerous domain boundaries.[41, 43–44] Apparently, increasing copper grain size can reduce the grain boundaries and densities of nucleation sites. Consequently, such reductions decrease the fraction of domain boundaries, which improves graphene structure. Thus, increasing grain size and crystal fraction can logically contribute to producing high-quality MLG, as shown in Figure 6(b). Therefore, controlling copper grain boundary is important in the RTT process for growing high-quality MLG.

Transport measurements based on a bottom-gate field-effect transistor configuration were also conducted to characterize the grown MLG. The typical transfer characteristic is shown in **Figure 7**(a). By applying the widely used fitting device model that combines minimum carrier density at the Dirac point and quantum capacitances, carrier mobility can be extracted from the transfer curve. The extracted carrier mobilities of electrons and holes for the device are ~2,600 $cm^2V^{-1}s^{-1}$ and ~3,000 $cm^2V^{-1}s^{-1}$, respectively, with a residual carrier density at the Dirac point of ~$1\times10^{12}$ $cm^{-2}$ at room temperature. The high carrier mobilities further support the low defect density in the grown MLG. The high quality of the grown MLG was also confirmed by the Shubnikov–de Haas oscillation (SdHO) measurement.[45] Figure 7(b) shows the experimental data of SdHO at a gate voltage of −30 V with a vertically applied magnetic field that varies from 0 T to 8 T. The SdHO values are indicated at each peak. These SdHO peaks are uniformly spaced as a function of $1/B$ as shown in the inset in Figure 7(b). The intercept extrapolated by the linearly fitting curve reveals the origin of the filling factor, $N = 1/2$, which indicates a Berry's phase of $\pi$, that is, the characteristic of the standard half-integer quantum oscillations of MLG.[1]

In summary, we present a simple and effective transfer-free RTT process for the fast growth of high-quality, large-scale MLG on $SiO_2$ substrates using solid carbon sources. In the RTT process, the quality of the thin copper layer inserted between a solid carbon layer and a $SiO_2$/Si substrate is a critical factor in controlling graphene quality. Our graphene-based devices exhibit satisfactory electrical properties, including a promising carrier mobility of 3,000 $cm^2V^{-1}s^{-1}$ and standard half-integer quantum oscillations. This work provides a controllable, effective, and economical transfer-free route to obtain high-quality, large-scale MLG for practical applications.

**Experimental Section**

*Materials*: We used copper films with a thickness of ~500 nm sputtered on a $SiO_2$/Si substrate. The copper films were sputtered under a base pressure of ~2×10$^{-6}$ Torr. The copper coating rate was below 0.1 A/s to ensure the high quality of copper films. The copper films were annealed at 1000 °C in Ar/$H_2$ (1:1) gas atmosphere under ambient-pressure for 6 hours. The cooling rate was 0.01 °C/s. PMMA (MICROCHEM 950 A2) and sucrose ($C_{12}H_{22}O_{11}$) were used as the solid carbon sources. The PMMA films spin-coated onto the copper film was about 100 nm thick. While the thickness of $C_{12}H_{22}O_{11}$ dilute layer on copper film was around 500 nm.

*RTT process*: The $SiO_2$/Si substrates containing annealed copper films and solid carbon sources on top were placed in the center of the processing chamber on a 4-inch Si substrate. The sample was then heated to the growth temperature with a heating rate of 160 °C/s protected by Argon gas. After staying at the growth temperature for 4-8 minutes, the system was cooled down to room temperature in about 30 minutes.

*Electronic Measurement*: The electronic properties of the as-grown graphene were evaluated based on the bottom-gate field-effect-transistor configuration. The metal electrodes were prepared by standard electron-beam lithography in Raith e_LiNE system and then electron-beam thermal evaporation (Ti/Au 5/45 nm). Transport measurements were carried out using standard lock-in technique (SR830) with a current source provided by Keithley (model 6221). The measured data are shown in Figure 7 and the extracted carrier mobility of electrons and holes for this device is ~2,600 $cm^2V^{-1}s^{-1}$ and ~3,000 $cm^2V^{-1}s^{-1}$, respectively (the residual carrier concentration at the charge neutrality point is ~$1\times10^{12}$ $cm^{-2}$).

*DFT calculation*: CASTEP code is used with an ultrasoft pseudopotential and PBE-style generalized gradient approximation.[46-47] Transition state search is also performed with CASTEP.[48] The cutoff energy is set to 400eV. A 108 supercell cell is used for bulk Cu and a 216-atom supercell is used for edge dislocation model. A vacuum layer is inserted for edge dislocation to relax the strain. A residual force of 0.02 eV/A is used as convergence criteria for geometry optimization. The similar parameter has been used in our previous study on metal catalyst.[49]

**Acknowledgements**
Financial supports from the Research Grants Council of Hong Kong (Project Nos. N_HKUST613/12 and 604112), NSF of China (Project No. 61274140), CRF project HKU9/CRF/13G and technical support of the Raith-HKUST Nanotechnology Laboratory for the electron-beam lithography facility (Project No. SEG HKUST08) are hereby acknowledged.

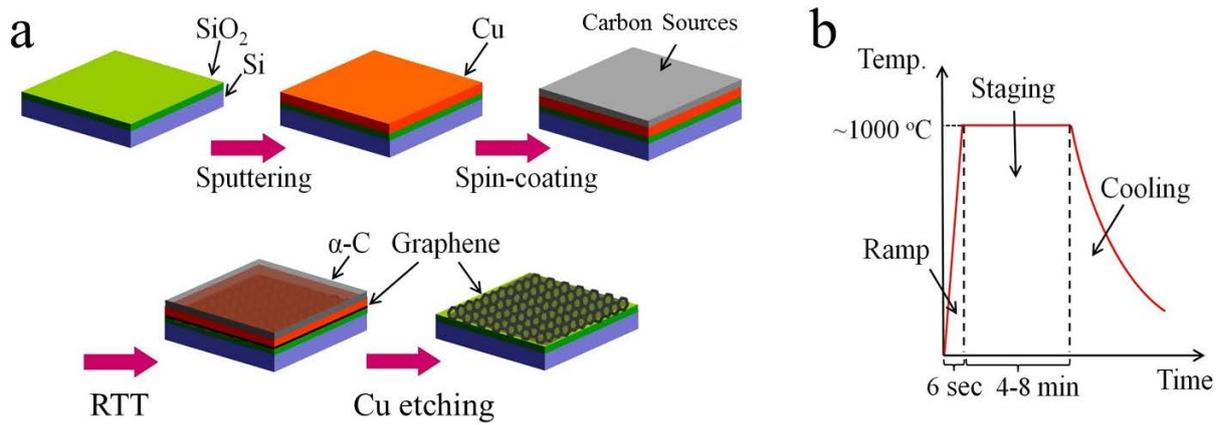

**Figure 1.** Schematic illustration for the transfer-free growth of graphene on insulating substrates. a) Copper films are sputtered on a $SiO_2$/Si substrate with the thickness of 500 nm. After annealing the copper films at 950 °C, solid carbon source such as PMMA or sucrose is spin coated on top of the film. The RTT process is carried out with the heating rate of 160 °C /s. After keeping at the staging temperature (800-1000 °C) for 4-8 minutes, the sample is cooled down to room temperature in about 30 minutes. Afterwards, on top of the copper film the amorphous carbon forms. While at the Cu-$SiO_2$ interfaces, there is large area monolayer graphene. The Cu layer together with α-C is finally dissolved away in a 0.1M $FeCl_3$ solution. And monolayer graphene formed directly on the $SiO_2$/Si wafer is obtained without any further process. b) the heating, staging and cooling processes.

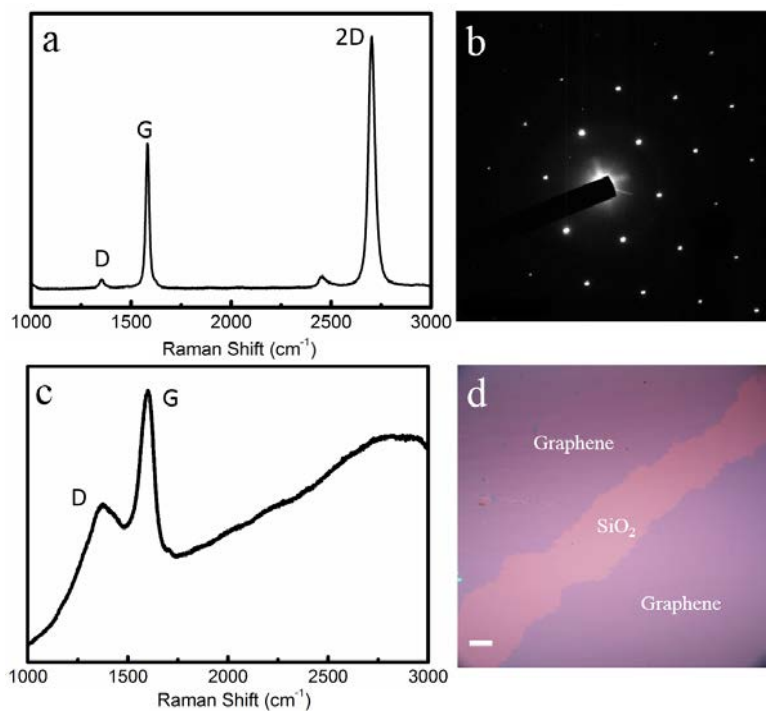

**Figure 2.** Characterization of as-grown graphene samples. a) Raman spectrum of a Randomly selected area of as-grown graphene shows the D, G and 2D peaks. b) The TEM diffraction pattern of as-grown graphene sample, displaying the typical hexagonal crystalline structure of graphene. c) Raman spectrum of as-grown amorphous carbon on top of the copper film. d) an optical image of the as-grown graphene on a SiO$_2$ substrate. The scale bar is 100 μm.

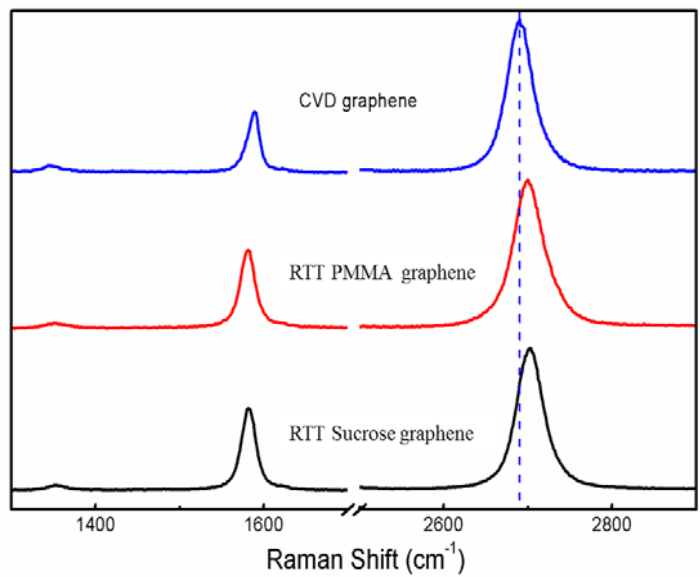

**Figure 3.** Raman spectrum of as-grown graphene film grown on top of the copper film through regular CVD method using methane and hydrogen as carbon source (in blue), and Raman spectra of PMMA/sucrose-derived graphene films grown on $SiO_2$/Si through RTT method at 1000 °C, respectively(in red/black).

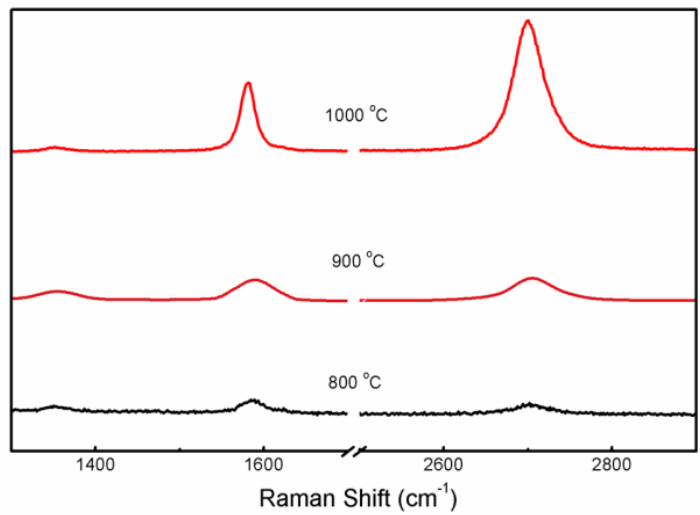

**Figure 4.** Raman spectra of PMMA-derived graphene films grown at different temperatures through RTT method.

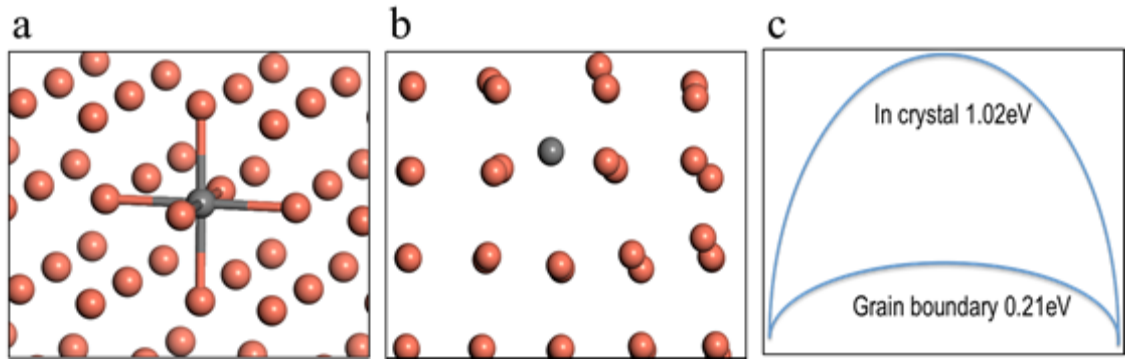

**Figure 5.** a) The relaxed octahedral interstitial site in Cu crystal. b) The relaxed interstitial site on an edge dislocation. c) The diffusion barrier along Cu(111) direction in a) and the edge dislocation direction in b).

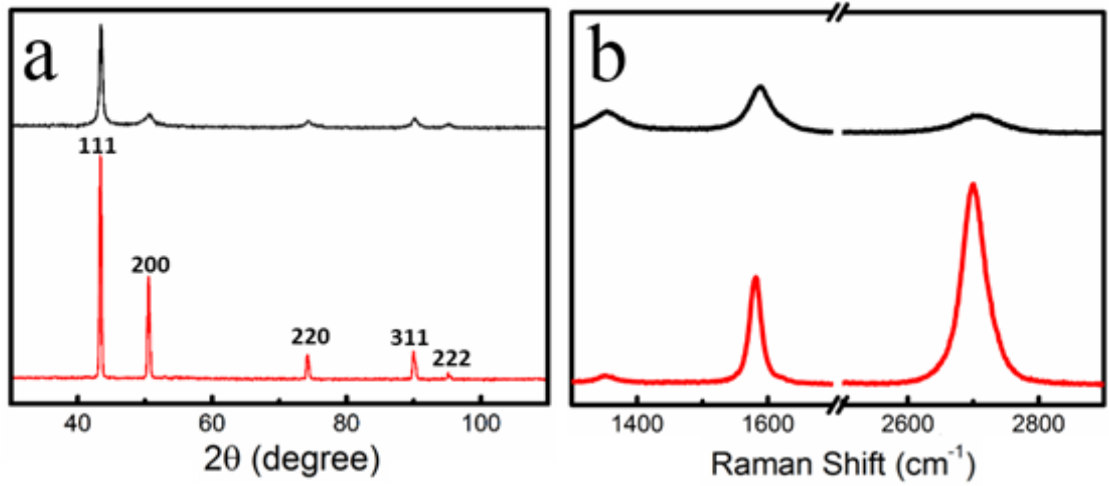

**Figure 6.** a) X-ray diffraction spectra of the sputtered copper films before (in black) and after (in red) annealing. B) Raman spectra of as-grown graphene films using the non-annealed/annealed (black/red) copper film.

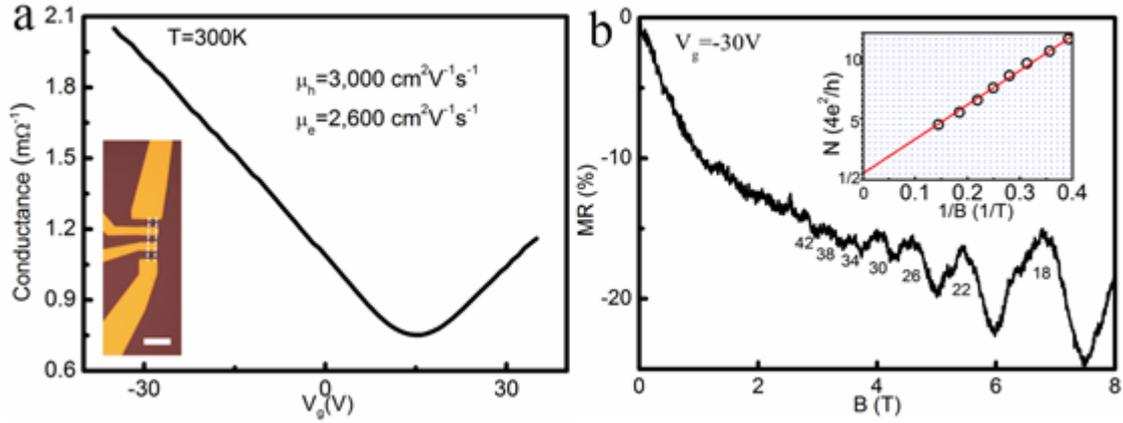

**Figure 7.** a) Transport measurement of PMMA-derived MLG by RTT method at T=300 K. The inset is an optical image of the FET device with the scale bar of 20 micron. b) Quantum oscillations in the corresponding MLG. SdHO at constant gate voltage $V_g$=-30 V as a function of magnetic field B is shown. The inset shows the SdHO peaks that are uniformly spaced as a function of 1/B.